# Evaluating authorship disambiguation quality through anomaly analysis on researchers' career transition

Huaxia Zhou[1†] and Mengyi Sun[2†]

[1]Department of Electrical and Computer Engineering, Northwestern University, Evanston, IL. USA.
[2]The Simons Center for Quantitative Biology, Cold Spring Harbor Laboratory, Cold Spring Harbor, New York, USA
[†]These authors contributed equally to this work.

**Abstract**

Authorship disambiguation is crucial for advancing studies in science of science. However, assessing the quality of authorship disambiguation in large-scale databases remains challenging, since it is difficult to manually curate a gold-standard dataset that contains disambiguated authors. Through estimating the timing of when 5.8 million biomedical researchers became independent Principal Investigators (PIs) with authorship metadata extracted from the OpenAlex— the largest open-source bibliometric database—we unexpectedly discovered an anomaly: over 60% of researchers appeared as the last authors in their first career year. We hypothesized that this improbable finding results from poor name disambiguation, suggesting that such an anomaly may serve as an indicator of low-quality authorship disambiguation. Our findings indicated that authors who lack affiliation information, which makes it more difficult to disambiguate, were far more likely to exhibit this anomaly compared to those who included their affiliation information. In contrast, authors with Open Researcher and Contributor ID (ORCID)—expected to have higher quality disambiguation—showed significantly lower anomaly rates. We further applied this approach to examine the authorship disambiguation quality by gender over time, and we found that the quality of disambiguation for female authors was lower than that for male authors before 2010, suggesting that gender disparity findings based on pre-2010 data may require careful reexamination. Our results provide a framework for systematically evaluating authorship disambiguation quality in various contexts, facilitating future improvements in efforts to authorship disambiguation.

# Introduction

Large-scale bibliometric databases form the backbone of quantitative studies in the science of science[1]. These comprehensive repositories—collecting scholarly works and associated metadata such as authorship, affiliations, and disciplinary classifications—enable investigations into research productivity[2], collaboration networks[3], and the impact of scholarly works[4]. However, the precision and reliability of such analyses critically depend on the quality of authorship disambiguation. Variations in spelling, cultural naming conventions, and incomplete metadata could cause multiple distinct individuals to be conflated under a single author profile or, conversely, a single individual to be split across multiple profiles[5].

Authorship disambiguation errors can bias findings about researcher careers, collaboration networks, and institutional or disciplinary dynamics. Although different algorithms and tools have emerged to address these challenges[6–8], evaluating the quality of disambiguation at scale remains difficult. Manual curation of large and representative datasets as a gold standard is rarely feasible due to the challenge of unbiased sampling of researchers with diverse demographic backgrounds. For example, ORCID[9], a self-reported authorship database that has been widely used as the gold standard for authorship disambiguation algorithms, suffers from highly uneven registration rate across countries[10].

In this study, we identify a novel, data-driven signature of poor authorship disambiguation quality. Our analysis centered on typical career progression in science, the transition from an apprenticeship to independent Principal Investigator (PI) status, often estimated from the first instance as the last author on a publication. We employed "last author analysis" approach[11], collecting relevant metadata from the OpenAlex[12]—the largest open-source bibliometric database—to estimate when approximately 5.8 million biomedical researchers became independent PIs.

Contrary to expectations, we found a striking irregularity: more than 60% of these researchers appeared as the last authors during their career debut years of publishing. Such a pattern is almost improbable under practical career trajectories, and its persistence across multiple years strongly suggests a systematic issue in authorship disambiguation. Further analyses confirmed that this anomaly was more common among authors lacking essential metadata—such as affiliation information or an ORCID—making them more challenging to reliably identify. Moreover, extending our examination to the dimension in gender revealed that data on female authors—particularly before 2010—were more frequently affected by these anomalies, raising concerns about the validity of historical analyses on gender disparities.

To address this striking anomaly finding, our study provides a framework for diagnosing and quantifying the authorshup disambiguation quality in bibliometric databases. Through extending the reliance on manually curated datasets, our approach identifies systematic biases and provides actionable insights for improving metadata quality at author level. Given the widespread use of bibliometric analysis in shaping science policy, informing funding decisions, and guiding scholarly evaluation, ensuring the accuracy and reliability of foundational data is crucial. Our findings contribute to advance the robustness and fairness of large-scale empirical research in the science of science.

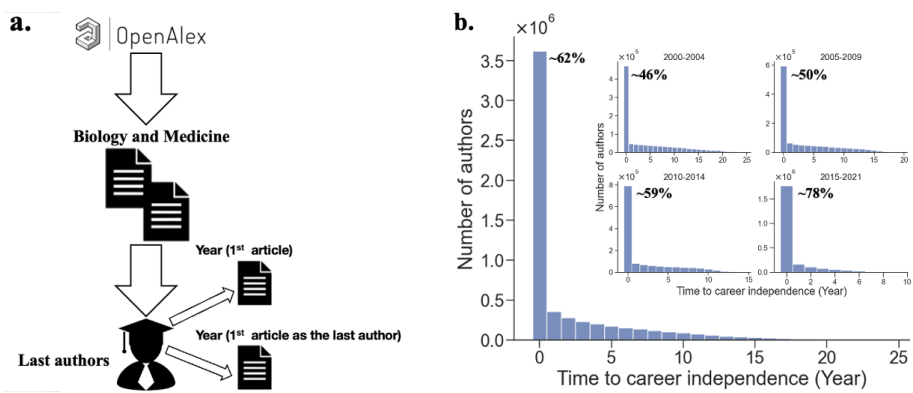

Figure 1: **The flow of execution for estimate the time to career independence for researchers in biomedical fields shows a stiking anomaly. a.** In the authorship conventions of academic publishing—particularly in biomedical research—the last author is often recognized as the Principal Investigator (PI) or senior researcher, indicating a level of career independence. The number of years it takes to achieve career independence can be estimated by calculating the difference between the career debut year—the year of a scholar's first publication—and the career independence year—the year of their first publication as the last author. This approach provides a measurable estimation of their progression within the career trajectory. **b.** An unusual anomaly in the estimation of career independence within the biomedical field has persisted over multiple years. Specifically, over 62% of authors published their first last-author paper in the same year they debuted with their first publication. This unexpected pattern of immediate career independence remained consistent across cohorts that began in different years, suggesting a systemic issue rather than isolated incidents.

# Results

## The Abnormally high rate of obtaining career independence in the career debut year

To perform "last author analysis", we processed the snapshot of OpenAlex database to obtain the career trajectories of biomedical researchers who began their careers after 1999 (see **Materials and Methods**). Specifically, we define a biomedical researcher as an author who publishes journal articles classified under either Biology or Medicine at Level 0 Concept within the OpenAlex taxonomy. Since in biomedical disciplines career independence is typically indicated by being the last author on a journal article, we then narrowed our focus on scholars publishing in these fields.

We defined an individual's career inception as the year of their earliest publication in the database and focused our analysis on authors with first publications after 1999, following the tradition of scientific career analysis[13, 14]. This time frame mitigates the issue of missing authorship records stemming from PubMed's former policy of capping the number of listed authors[15]—a limitation that affected OpenAlex, where PubMed is a primary source for biomedical literature[12].

In total, we obtained the career transition of 5,803,782 authors from their journal publication records. We retrieved 48,000,188 journal articles associated with this cohort of authors. The career transition to secure a tenure-track position in the biomedical research field is notably longer than in other disciplines, primarily due to the bottleneck created by the 'postdoc queue'[16]. This phenomenon reflects the extended periods post-Ph.D. that researchers often spend in postdoctoral positions before obtaining tenure-track roles, navigating a competitive and saturated job market. Focusing on this cohort of authors who had published at least one paper as the last author, we specifically calculated the duration from the inception of their career to the publication of their first last-author paper (See Fig 1a). This estimation is under the premise that last-author publications signify the attainment of a tenure-track position.

Our findings were startling: approximately 62% of these authors published their first paper at last author positon in the same year they debuted with their first publication (see Fig 1b). This improbable result suggests that a significant proportion of authors are achieving what is traditionally considered a marker of career independence at the very onset of their careers, which contradicts the typical progression in academic careers. Furthermore, this trend remained consistent across cohorts starting in different years (inset of Fig 1b), indicating that this is a systemic issue rather than an anomaly specific to a particular time period.

## Poor authorship disambiguation is the cause of the anomaly

This observed pattern raises concerns about potential errors in the authorship name disambiguation process. If the abnormally high rate of becoming a PI in the first year result from such errors, these anomalies should be more frequent in scenarios where authorship disambiguation is particularly challenging.

To investigate this, we analyzed two scenarios: we first compared authors with and without affiliation information, hypothesizing that missing affiliation data—a critical feature for disambiguation—would result in higher anomaly rates. Next, we examined authors with and without ORCID identifiers. Since ORCID relies on self-reporting by authors who have full knowledge of their publication records, it is widely regarded as a gold standard for disambiguation. We hypothesized that authors with ORCID identifiers would show lower anomaly rates[9].

To test these hypotheses, we classified those authors based on the presence or absence of affiliation and ORCID metadata. We calculated the percentage of authors who achieved last authorship in their debut year among those who became last authors within five years of their career start, focusing on individuals who debuted between 2000 and 2018 to address right-censoring.

As shown in Fig. 2a, authors missing affiliation metadata exhibited consistently high anomaly rates, exceeding 80% across the 18-year period. This suggests that many were erroneously recorded as last authors in their career debut year, an implausible outcome indicative of disambiguation errors. In contrast, authors with complete affiliation metadata had lower and more consistent anomaly rates of approximately 60%.

We observed a similar pattern for authors with ORCID data. Authors with ORCID identifiers showed an anomaly rate of about 40% over the 18-year period, significantly lower than the 70% observed among authors without ORCID (Fig. 2b).

Together, these findings strongly support our hypothesis that errors in authorship disambiguation highly contribute to the observed anomalies.

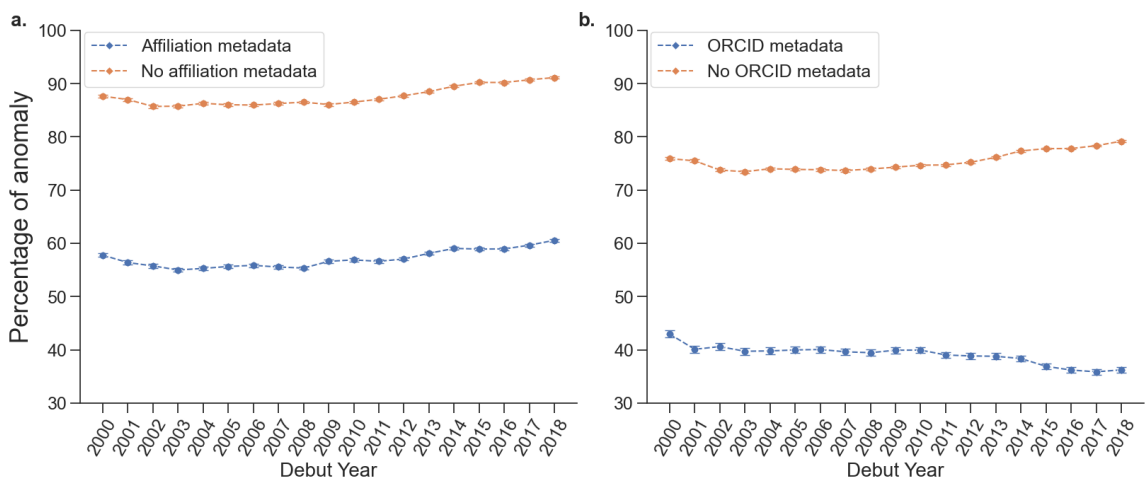

Figure 2: **Authors with essential identifiers—affiliation and ORCID—exhibit lower percentages of anomaly. a.** Percentage of authors who achieved last authorship within five years of their career debut, comparing those with and without country metadata. Authors lacking country metadata consistently show an anomaly rate exceeding 80% over 18 years, while those with country metadata exhibit a lower and more stable anomaly rate around 60%. **b.** Percentage of authors who achieved last authorship within five years of their career debut, comparing those with and without ORCID identifiers. Authors without ORCID data have an anomaly rate above 70%, whereas authors with ORCID identifiers show a significantly reduced anomaly rate of about 40%.

## Using career anomaly as a warning signal for poor authorship disambiguation quality

Our findings on the relationship between anomaly rate and authorship metadata suggest that the anomaly rate is highly sensitive to disambiguation errors. Consequently, we propose that the anomaly rate can serve as a reliable warning signal for detecting poor authorship disambiguation quality in bibliometric databases used by biomedical researchers. We believe that identifying contexts contributing to poor name disambiguation is key to developing more effective disambiguation methods. This signal also acts as a reliability check for group comparisons, revealing potential data biases.

As an example, we investigated whether authorship disambiguation quality differs between male and female authors, given the significant focus on academic outcome disparities between these genders[17, 18]. To conduct this analysis, we first categorized authors into male and female groups using a name-based gender classification model[19]. We then followed a procedure similar to our earlier analysis of the relationship between anomaly rate and authorship metadata.

Our results reveal minor yet statistically significant discrepancies in authorship disambiguation quality between male and female authors. Specifically, female authors tend to exhibit a slightly higher anomaly rate, indicating a greater frequency of authorship disambiguation errors (Fig. 3a). This pattern is most pronounced before 2008 for all authors but becomes negligible or even reversed after that period (Fig. 3a). When we further segmented authors based on the continent of their affiliation, we observed similar patterns (Figs. 3,b-d), though with varying degrees of severity. These findings suggest that differences in authorship disambiguation quality between male and female authors may exist, warranting a re-evaluation of some gender-based bibliometric analyses to assess whether this differential quality introduces bias into their conclusions.

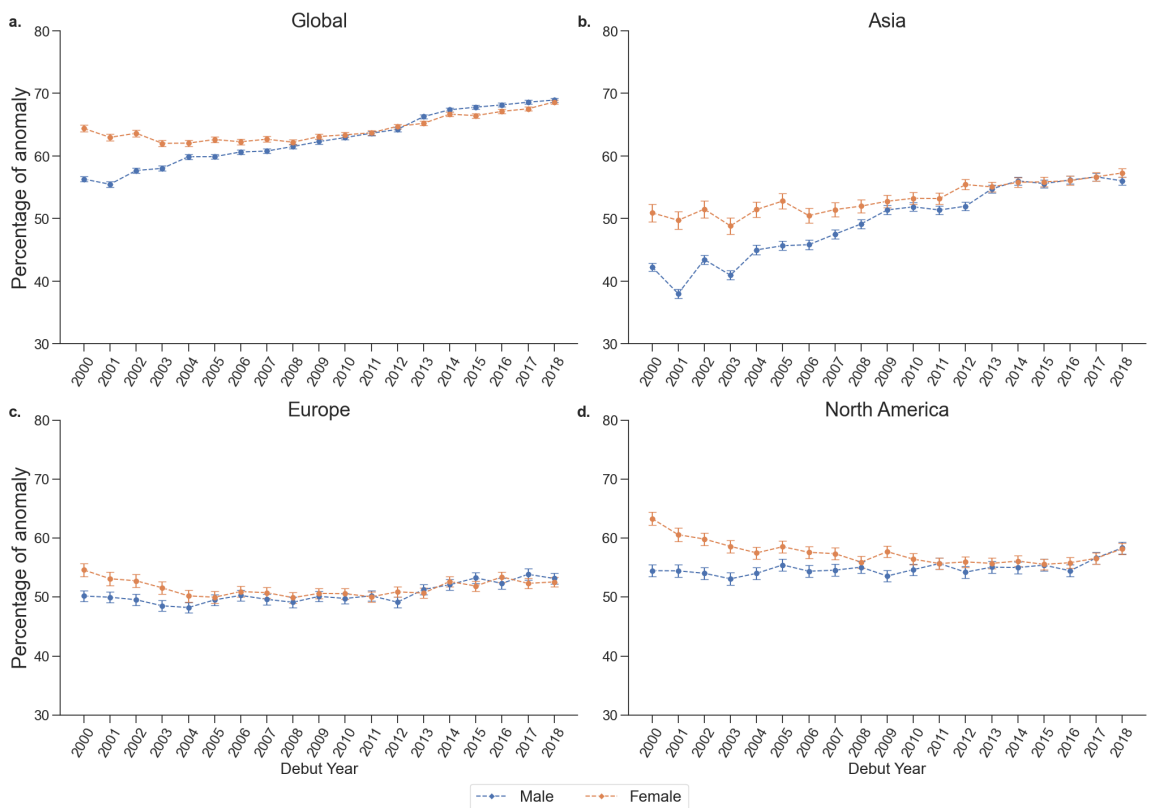

Figure 3: **Discrepancy in anomaly occured between male and female authors before 2008. a.** Percentage of male and female authors achieving last authorship within five years of their career debut, highlighting the discrepancy before 2008. Female authors exhibit a higher percentage of anomalies. **b.** Percentage of male and female authors affiliated with Asian institutions achieving last authorship within five years of their career debut, showing the discrepancy before 2013. **c.** Percentage of male and female authors affiliated with European institutions achieving last authorship within five years of their career debut, showing the discrepancy before 2005. **d.** Percentage of male and female authors affiliated with North American institutions achieving last authorship within five years of their career debut, showing the discrepancy before 2011.

## Discussion

In this study, we conducted "last author analysis" to 5.8 million authors in the biomedical field, using data from the OpenAlex database. Our analysis uncovered a striking anomaly: 62% of biomedical researchers achieved last authorship—an indicator of career independence—within their career debut year. We further demonstrated that this anomaly is sensitive to the quality of authorship disambiguation. By analyzing the relationship between the anomaly rate and authorship metadata, such as affiliation and ORCID information, we revealed that poor disambiguation quality plays a significant role. Finally, we proposed using this anomaly as a warning signal to investigate contexts contributing to poor authorship disambiguation. Our findings highlight differential disambiguation quality between male and female authors and underscore the potential need to revisit gender-based career outcome patterns derived from bibliometric databases.

Our study has several limitations. First, our analysis relies on the norm in biomedical research, where last authors typically serve as corresponding authors[20]. This convention might not fully apply to other fields. Even in biomedical fields, some journals may not list the corresponding author at the last author position. Future research should explore the extent to which our proposed warning signal can be generalized to other disciplines and check with available data on corresponding author rather than last author. Secondly, our analysis focused solely on the OpenAlex dataset, leaving open the question of whether similar issues exist in other databases, particularly proprietary databases (i.e. Scopus[21], Dimensions[22], and Web of Science[23]). Comparative studies across multiple databases are needed to check whether our finding is a universal anomaly. Finally, our findings on gender-based disparities in authorship disambiguation quality relied on name-based gender inference, which is often imprecise[24]. However, since previous research on gender disparities has mostly used name-based methods[18, 25, 26], our results suggest broader limitations inherent in such analyses.

Our research opens several possibilities for future exploration. The identified anomaly could significantly affect analyses of early-career scientists' trajectories, making it important to develop methods that correct for this issue using advanced authorship disambiguation techniques or robust comparisons under noisy publication records. Furthermore, while we observed gender-based differences in disambiguation quality, the underlying causes remain unclear. Comprehensive studies are needed to uncover and address these factors to better document gender disparities in academia. Finally, identifying additional scalable warning signals for assessing authorship disambiguation quality and exploring associated contexts represent exciting directions for future research. We believe our findings will inspire further rigorous investigations in these areas.

# Materials and Methods

## OpenAlex data, preprocessing and last-author analysis

We downloaded the OpenAlex data (Snapshot released on April.19 2024) from Amazon S3 storage and maintained the snapshot on the high-performance computing cluster. From this dataset, we selected authors based on the following two criteria:

1. Authors published journal articles fall within the fields of Biology or Medicine (Concept IDs are: https://openalex.org/C86803240 or https://openalex.org/C71924100).

2. All selected publications are non-retracted journal articles. We filtered out para-texts or retracted articles.

This selection yielded 66,165,323 journal article publications (authored by 37,019,107 authors). We then retrieved all publications for these authors and determined their authorship positions based on the author order in the database. An author is defined as the last author if the author position is the last in the publication, and the author position feature can be directly accessed in OpenAlex. We filtered for authors who served as the last author in at least one publication, resulting in 5,806,208 authors who published their first publication after 1999. For this cohort of authors, we calculated the time to career independence as the duration between their first publication and their first publication as the last author.

### Authorship Metadata

Although about 95% of journal articles can be connected to authorship information in the OpenAlex (195,447,128 out of 205,389,471 journal articles can be linked to "AuthorId" and "AuthorPosition"). However, authors may not have their affiliation information available in OpenAlex. From the authors entity in the OpenAlex database, we could retrieve author metadata, including authors' affiliated institutions (We used the last known institution as an estimation of the author's most recent affiliation). In total, about 37% of authors have affiliations information available (33,197,783 out of 90,556,187 authors can be linked to "AffiliationId").

### Gender Prediction

To predict the gender of the selected authors, we extracted their first and last names and applied the "nomquamgender" package[19]. This package used name-based gender classification to estimate the likelihood that an individual has been structurally gendered female or male, providing probabilities rather than definitive classifications. By focusing on names as a narrow data stream, the method reflects only a shadow of the gendering process, aligning with the conceptual framework of "nomquamgender" as "name rather than gender". Thus, this approach ensures that classifications are interpreted as probabilities of gendering rather than traditional gender labels, consistent with best practices.

We classified authors based on the probabilities provided by the "nomquamgender" package, which estimates the likelihood that a name corresponds to an individual who has been structurally gendered female ($p(gf)$). Specifically, we categorized individuals as male if $p(gf) <= 0.2$, and as female if $p(gf) >= 0.8$. This threshold-based approach ensures a high level of confidence in the classifications by focusing on names strongly associated with one gender. Authors with probabilities falling between 0.2 and 0.8 were not classified into either category, reflecting the inherent uncertainty for such cases and maintaining the methodological rigor of the analysis.

# Acknowledgments

We want to thank D.Acuña and L.A.N.Amaral for insightful discussions during the preparation of the project, and specially thanks to D.Acuña for providing us with the 2023 version of OpenAlex snapshot data to perform the our analysis. We also want to express our gratitude to the participants at ICSSI3 who provided us with suggestive feedback on this study.